\documentclass[lettersize,journal]{IEEEtran}
\usepackage{amsmath,amsfonts}
\usepackage{algorithmic}
\usepackage{algorithm}
\usepackage{array}
\usepackage[caption=false,font=normalsize,labelfont=sf,textfont=sf]{subfig}
\usepackage{textcomp}
\usepackage{stfloats}
\usepackage{url}
\usepackage{verbatim}
\usepackage{graphicx}
\usepackage[numbers,sort&compress]{natbib}
\usepackage[utf8]{inputenc}
\usepackage{hyperref}
\usepackage{booktabs}
\usepackage{multirow} 
\usepackage{amssymb}
\usepackage{makecell}
\usepackage{pifont}
\usepackage[table,xcdraw]{xcolor}
\hypersetup{
    colorlinks=true,
    linkcolor=blue,
    citecolor=blue,
    urlcolor=blue,
    linkbordercolor={1 1 1},
    citebordercolor={1 1 1},
    urlbordercolor={1 1 1}
}
\begin{document}

\title{A Comprehensive Review of Protein Language Models}

\author{
    \IEEEauthorblockN{Lei Wang\textsuperscript{1†}, Xudong Li\textsuperscript{1†}, Han Zhang\textsuperscript{1†}, Jinyi Wang\textsuperscript{1}, Dingkang Jiang\textsuperscript{3}, Zhidong Xue\textsuperscript{1*}, and Yan Wang\textsuperscript{2*}} \\
    \IEEEauthorblockA{$^1$School of Software Engineering, Huazhong University of Science and Technology.} \\
    \IEEEauthorblockA{$^2$School of Life Science and Technology, Huazhong University of Science and Technology.} \\
    \IEEEauthorblockA{$^3$School of Computer Science and Technology, Huazhong University of Science and Technology.} \\
    \IEEEauthorblockA{* Correspondence author: yanw@hust.edu.cn (Y.W.), zdxue@hust.edu.cn (Z.X.)} \\
        \IEEEauthorblockA{† indicates equal contribution.} \\ 
}

\markboth{Journal of \LaTeX\ Class Files,~Vol.~14, No.~8, December~2024}%
{Shell \MakeLowercase{\textit{et al.}}: A Sample Article Using IEEEtran.cls for IEEE Journals}


\maketitle

\begin{abstract}
At the intersection of the rapidly growing biological data landscape and advancements in Natural Language Processing (NLP), protein language models (PLMs) have emerged as a transformative force in modern research. These models have achieved remarkable progress, highlighting the need for timely and comprehensive overviews. However, much of the existing literature focuses narrowly on specific domains, often missing a broader analysis of PLMs. This study provides a systematic review of PLMs from a macro perspective, covering key historical milestones and current mainstream trends. We focus on the models themselves and their evaluation metrics, exploring aspects such as model architectures, positional encoding, scaling laws, and datasets. In the evaluation section, we discuss benchmarks and downstream applications. To further support ongoing research, we introduce relevant mainstream tools. Lastly, we critically examine the key challenges and limitations in this rapidly evolving field.
\end{abstract}


\section{Introduction}
\begin{figure}
\centering
\includegraphics[scale=0.27]{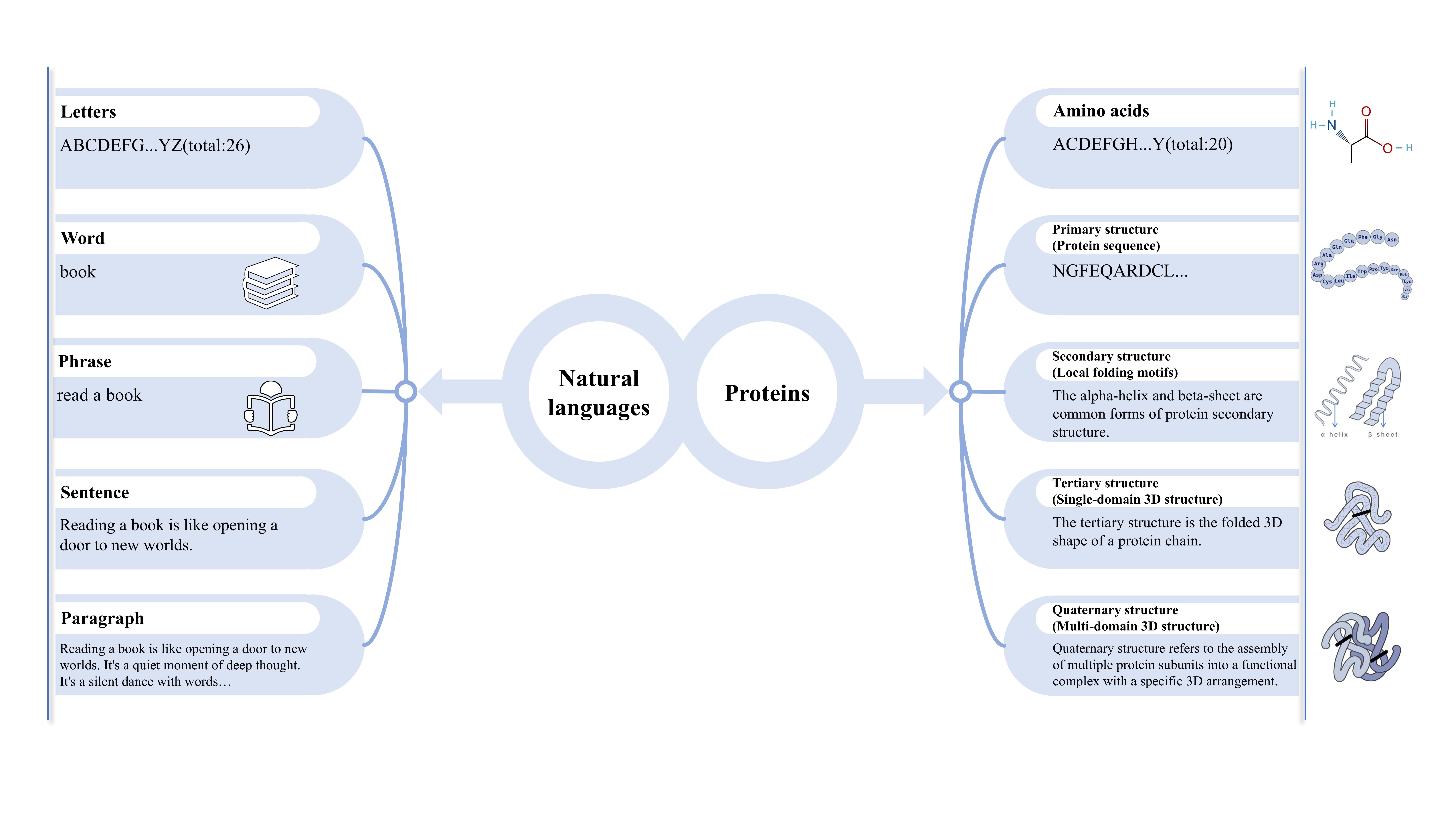}
\caption{The conceptual similarities and hierarchical structures observed in natural languages and proteins.}
\label{fig1} 
\end{figure}

\IEEEPARstart{I}{n} recent years, the success of the Transformer\cite{1transformer} in the field of NLP has attracted much attention. Due to the significant similarity between protein sequences and natural languages, NLP have been introduced into the field of protein studies. Like human language, protein sequences, composed of linear residue chains, can be naturally represented as strings of letters, with protein strings consisting of 20 common amino acids (AAs)(excluding unconventional and rare amino acids)\cite{187}. Although proteins show complex and variable three-dimensional(3D) structures that are closely related to the environment, it is still defined by the underlying amino-acid sequence\cite{2}. This means that, theoretically, sequence-based methods can learn all the information about proteins. Fig.\ref{fig1} illustrates the conceptual similarities and hierarchical structure observed in natural languages and proteins.

The rapid advancement of sequencing technologies has led to an exponential increase in label-free protein sequence data, which continues to grow. This extensive and diverse dataset provides a foundation for leveraging large-scale neural networks\cite{3,5esm1b}. By combining this data with Transformer-based architectures, a surge in protein language models has emerged. These models generate distributed embedded representations that encode semantic information about proteins through self-supervised learning, enabling significant advancements in various downstream tasks. These breakthroughs are increasingly matching or even surpassing the results of manual experiments, driving remarkable progress in the field\cite{5esm1b,188prottrans}. Given the transformative impact of PLMs, it is essential to provide a comprehensive summary of the advancements and work in this area.



\begin{figure*}[!t]
\centering
\includegraphics[width=\textwidth]{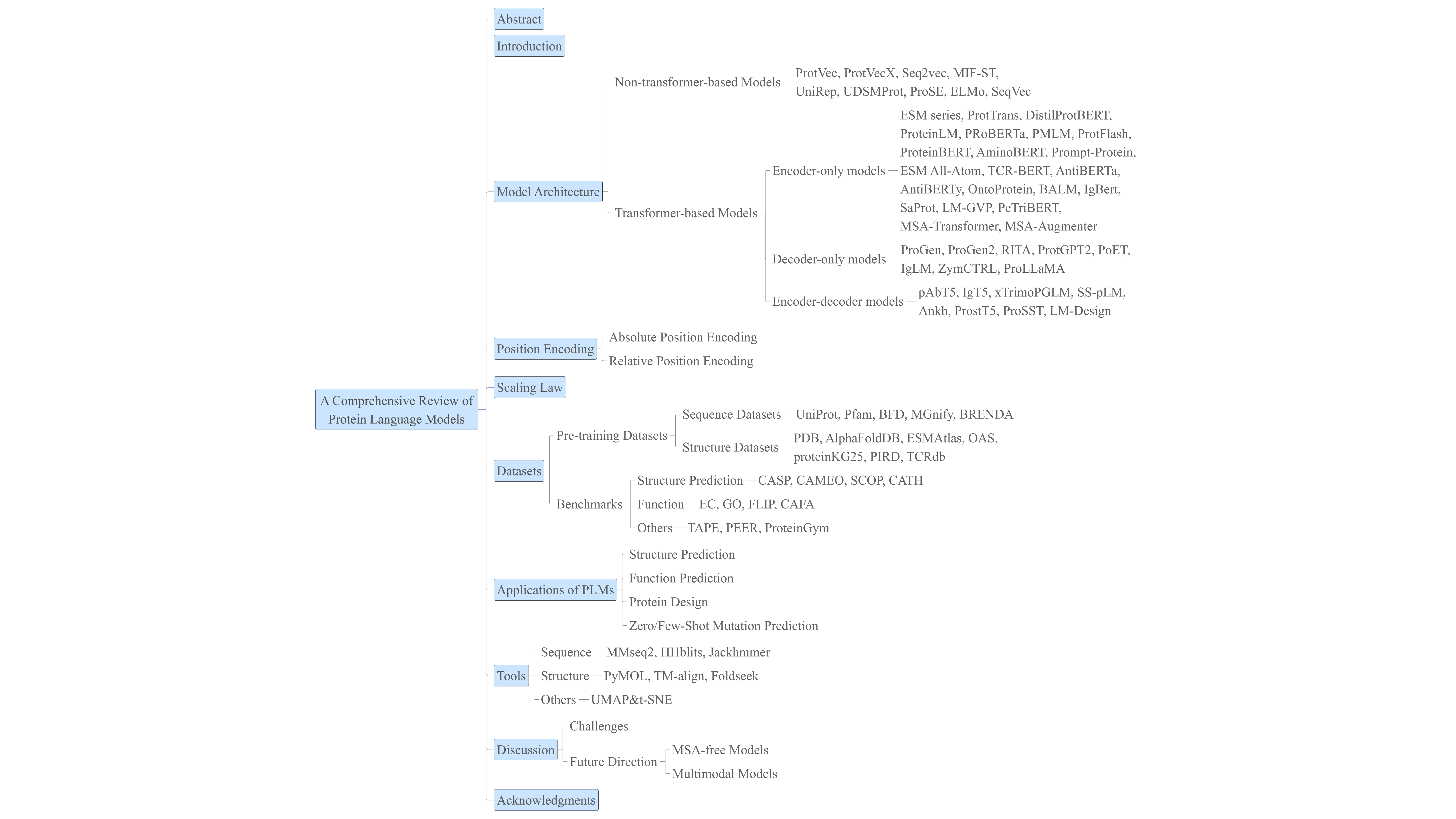} 
\caption{The framework diagram of the paper.}
\label{fig2} 
\end{figure*}

In this survey, we systematically review the technological advances in PLMs to fill this gap. As shown in Fig.\ref{fig2}, we summarize key factors that influence model performance, including model architecture, positional encoding, scaling laws, and pre-training datasets. Given that the evaluation of PLMs’ performance is dependent on downstream tasks\cite{5esm1b,9}, we will also discuss benchmarks and applications. We will intersperse representative models from various fields throughout the paper and provide a comparative discussion to offer a broader understanding of PLMs. Additionally, to facilitate research, we will introduce relevant mainstream tools. Finally, we analyze the major challenges currently faced and highlight key developmental milestones of PLMs over time, with a focus on identifying current mainstream trends.

We provide a comprehensive collection of resources related to major protein language models, datasets, and tools, along with links to their associated papers and code repositories, at \href{https://github.com/ISYSLAB-HUST/Protein-Language-Models}{https://github.com/ISYSLAB-HUST/Protein-Language-Models}. It is important to note that this survey is limited to language models. Although models based on other neural networks, such as CNNs and GNNs, have also played significant roles in the protein domain, they are not considered language models. Only when these networks are combined with language models (e.g., LM-GVP\cite{10lmgvp}, which is a joint model of GNNs and language models) are they included in our discussion.

\section{Model Architecture}
\label{Model Architecture}
In this section, we will comprehensively review common model architectures, including Non-Transformer-based and Transformer-based architectures, and present their practical applications through discussions of representative models.

\subsection{Non-transformer-based Models}

Non-Transformer-based models represent an early exploration of neural networks for protein feature representation\cite{5esm1b,11}, providing valuable experience for the subsequent successful application of Transformer-based architectures in the protein domain. Representative architectures include word2vec, doc2vec, CNN, GNN, and recurrent neural networks.Table \ref{Non-transformer-based models} summarizes Non-transformer-based models.

In 2013, Google’s research team introduced the word2vec technique\cite{12word2vec}, which uses shallow neural networks to map one-hot words to distributed word embeddings. ProtVec\cite{13protvec} was the first model to apply embedding methods to biological sequences which treats amino acid triplets as words and subsequently applies the skip-gram model of word2vec to the Swiss-Prot\cite{14uniprotkbswissprot} dataset, resulting in 100-dimensional protein embeddings. Due to its simplicity, ProtVec has inspired various extensions, primarily focusing on k-mer variations. ProtVecX\cite{15protvecx} is one such extension that uses word2vec to embed variable-length amino acid k-mers rather than fixed-length k-mers. Another extension, Seq2vec\cite{16seq2vec}, embeds entire protein sequences instead of k-mers, it utilizes doc2vec\cite{17doc2vec}, an extension of word2vec, which processes documents instead of individual words. MIF-ST\cite{18mifst} is a joint sequence-structure model composed of CNN\cite{19cnn} and GNN\cite{20gnn}, which enhances model performance by embeddings sequence representations derived from a CNN-based CARP-640M\cite{21carp} sequence model into a MIF graph network.

Common recurrent neural network structures include recurrent neural networks (RNN)\cite{23rnn} and long short-term memory networks (LSTM)\cite{24lstm}. UniRep\cite{25unirep} is an mLSTM (multiplicative LSTM)\cite{28mlstm} trained on the UniRef50\cite{26uniref,27uniref} protein sequence dataset to generate high-dimensional protein sequence representations. UDSMProt\cite{29udsmprot} adopts the AWD-LSTM\cite{30awdlstm} architecture, employing different dropout methods for accurate word-level language modeling. ProSE\cite{3} utilizes a three-layer bidirectional LSTM (BiLSTM) with skip connections and, through multi-task learning, can simultaneously process sequence, structure, and functional information. Allen Institute for Artificial Intelligence proposed a pre-trained language model based on bidirectional recurrent neural networks, ELMo (Embeddings from Language Models)\cite{31elmo}. ELMo represents a significant breakthrough, as it dynamically generates word embeddings based on context. SeqVec\cite{32seqvec} is the first model to use ELMo for amino acid representation based on entire protein sequences. This model trains on the entire UniRef50 dataset and predicts much faster than traditional methods relying on MSA. 

Recurrent neural networks address the limitations of early language models, such as N-grams\cite{34ngram}, which can only handle fixed-length sequences and long-distance dependencies. However, recurrent neural networks have limited parallelization capabilities, and they typically require a large amount of labeled data, making large-scale learning challenging\cite{23rnn,1transformer,35}. 

\subsection{Transformer-based Models}

Transformer has achieved revolutionary results in the processing of protein sequences \cite{37}. protein language models (PLMs) typically utilize heavily parameterized Transformer models as their base architecture\cite{38}. In this section, we will first introduce the classic Transformer\cite{1transformer}, BERT\cite{39bert}, and GPT\cite{40gpt} architectures, and focus on reviewing representative Transformer-based models. Transformer-based models can be classified into three types based on their specific architectures: encoder-only, decoder-only, and encoder-decoder. Table \ref{Encoder-only models},\ref{Decoder-only models},\ref{Encoder-decoder models} summarize Transformer models.

\subsubsection{Transformer}

\begin{figure}[h]
\centering
\includegraphics[width=4in, trim=0.9cm 0cm 5cm 0cm, clip]{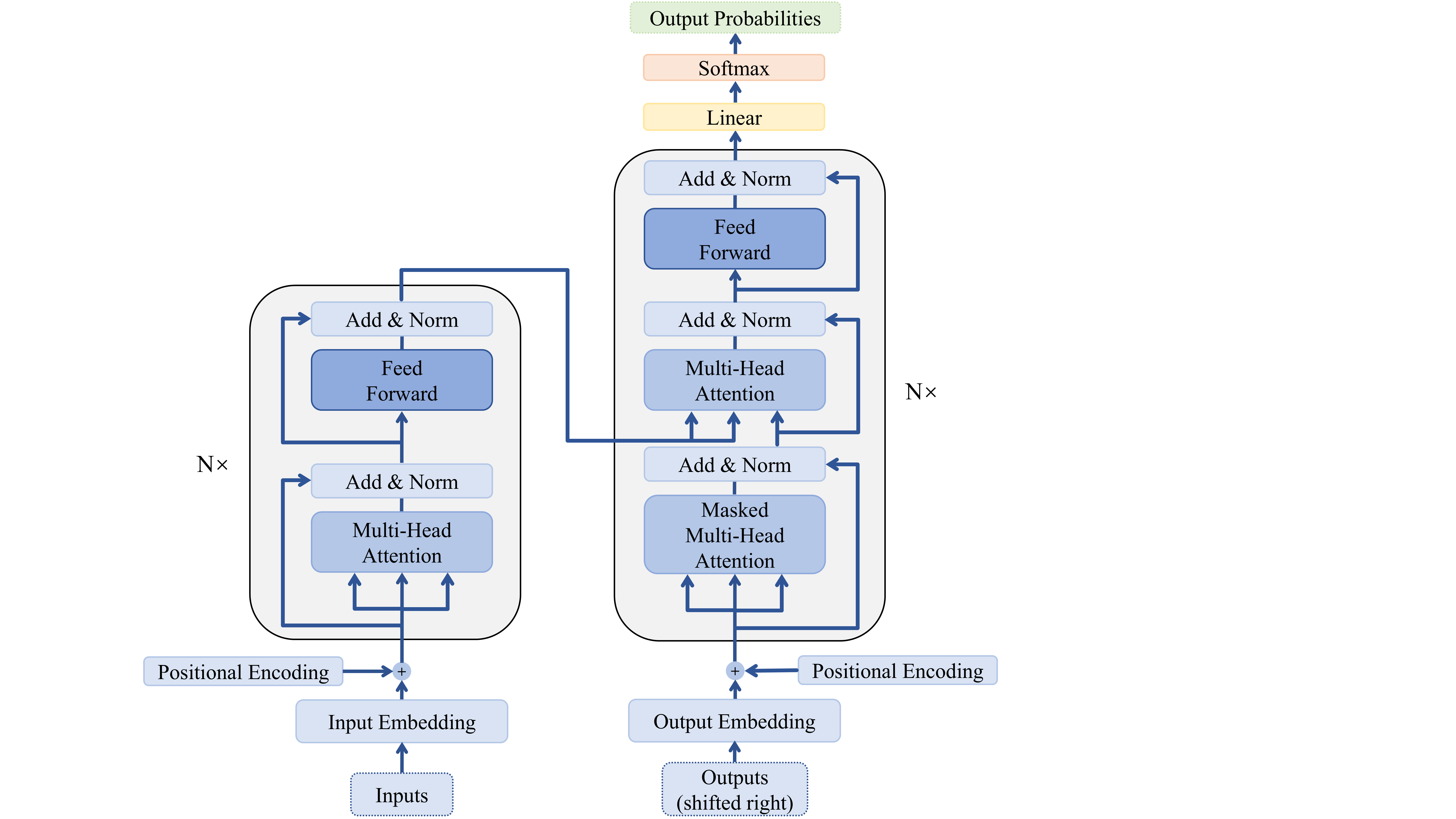}
\caption{The architecture of transformer.}
\label{fig3} 
\end{figure}

The complete Transformer\cite{1transformer} is an encoder-decoder architecture (Fig.\ref{fig3}), with BERT and GPT being improvements based on the Transformer encoder and decoder architectures, respectively.

Transformer is a groundbreaking deep learning model in natural language processing that processes input sequences in parallel through self-attention. The self-attention mechanism is a core component of Transformer, allowing the model to dynamically adjust the importance of information from different positions when processing each position in the input sequence, thus enabling it to better capture global dependencies. The introduction of multi-head attention grants the model the ability to capture different semantic relationships under various attention weights\cite{41}. Transformer consists of an encoder that extracts embedding representations of the input sequence, and a decoder that uses the encoder's output along with self-attention to generate the target sequence step-by-step, commonly used in sequence-to-sequence tasks. Fig.4 illustrates the architecture of the transformer model, highlighting its key components and the flow of information within the network. Transformer has made breakthrough advancements across various fields of natural language processing, laying the foundation for subsequent models like BERT and GPT\cite{42protflash}.

BERT\cite{39bert} is an advanced pre-trained language model based on Transformer encoder architecture, composed of multiple stacked encoder layers. It uses masked language modeling as the training objective to learn contextually relevant word representations, rather than just static word embeddings. A prominent feature of BERT is its bidirectional pre-training, allowing it to obtain more comprehensive language representations. These features endow BERT with exceptional semantic representation capabilities, resulting in outstanding performance in downstream tasks\cite{42protflash}. GPT(Generative Pre-trained Transformer)\cite{40gpt} is a pre-trained language model based on Transformer decoder architecture. GPT employs autoregressive generation and is the first pre-trained language model to successfully apply Transformer architecture to generative tasks. The core of the GPT model is its unidirectionality, processing sequence data from left to right, which makes it widely applicable for sequence generation tasks\cite{44}.

\begin{figure}[h]
\centering
\includegraphics[width=3.6in, trim=1.6cm 2cm 1cm 2cm, clip]{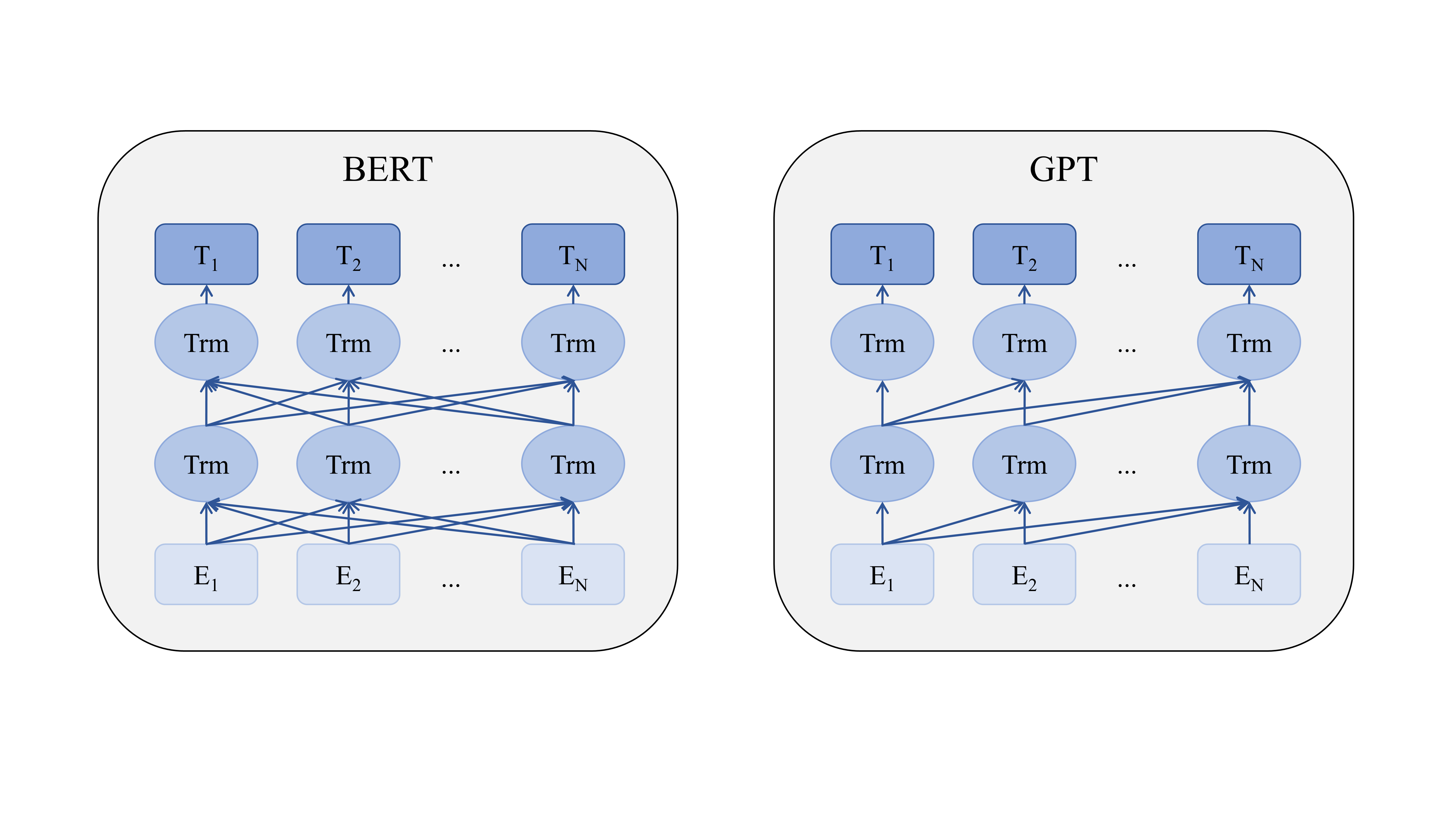}
\caption{The architecture of BERT and GPT.}
\label{fig4}
\end{figure}

Fig.\ref{fig4} presents the architectures of BERT and GPT, showcasing their key design elements. BERT adopts a bidirectional transformer architecture for deep contextual understanding, while GPT employs a unidirectional transformer optimized for autoregressive tasks.

\subsubsection{Encoder-only Models}

Encoder-only models, such as BERT \cite{39bert}, are commonly used to encode protein sequences into fixed-length vector representations, extracting features for various downstream tasks. Notable examples include Meta's ESM series \cite{5esm1b,7esm1v,47esm2,48esm3,ESMC}, which leverages large amounts of unlabeled protein sequence data, and ProtTrans \cite{188prottrans}, which trains various autoencoder models (e.g., BERT, ALBERT \cite{50albert}, ELECTRA \cite{51electra}) on extensive amino acid data from UniRef and BFD \cite{49bfd}. However, ProtTrans's large size makes it impractical for academic use, prompting the development of DistilProtBERT \cite{53distilprotbert}, a more efficient version with reduced computational requirements.

Several models have introduced novel improvements to BERT-like architectures. ProteinLM \cite{54proteinlm} removes the Next Sentence Prediction (NSP) task, focusing on Masked Language Modeling (MLM). PRoBERTa \cite{55proberta}, based on RoBERTa \cite{56roberta}, enhances dynamic masking and batch size via the LAMB \cite{57lamb} optimizer. PMLM \cite{60pmlm} uses a pairwise MLM that captures co-evolutionary data without Multiple Sequence Alignments (MSA). ProtFlash \cite{42protflash} reduces model complexity with a hybrid block attention mechanism. ProteinBERT \cite{58proteinbert} incorporates a global attention mechanism and multitask learning, while AminoBERT \cite{59aminobert} introduces new training objectives like masking consecutive residues and "block shuffling." PromptProtein \cite{75promptprotein} uses prompt-guided pre-training, and ESM All-Atom \cite{62esmallatom} unifies modeling at atomic and residue scales.

Antibody-specific models also exist, such as TCR-BERT \cite{63tcrbert}, which reduces positional encoding to 64 for shorter TCR sequences, and AntiBERTa \cite{64antiberta}, which predicts paratope positions. AntiBERTy \cite{65antiberty} uses the Multiple Instance Learning (MIL) framework for identifying antigen-binding sites, while OntoProtein \cite{66ontoprotein} integrates Gene Ontology knowledge into protein pre-training. BALM \cite{BALM} incorporates antibody-aware posi tional information into the position embedding, while IgBert \cite{IgBert} fine-tuned its unpaired version to learn cross-chain features.

\begin{figure}[h]
\centering
\includegraphics[width=3.5in, trim=0.9cm 4cm 0cm 0cm, clip]{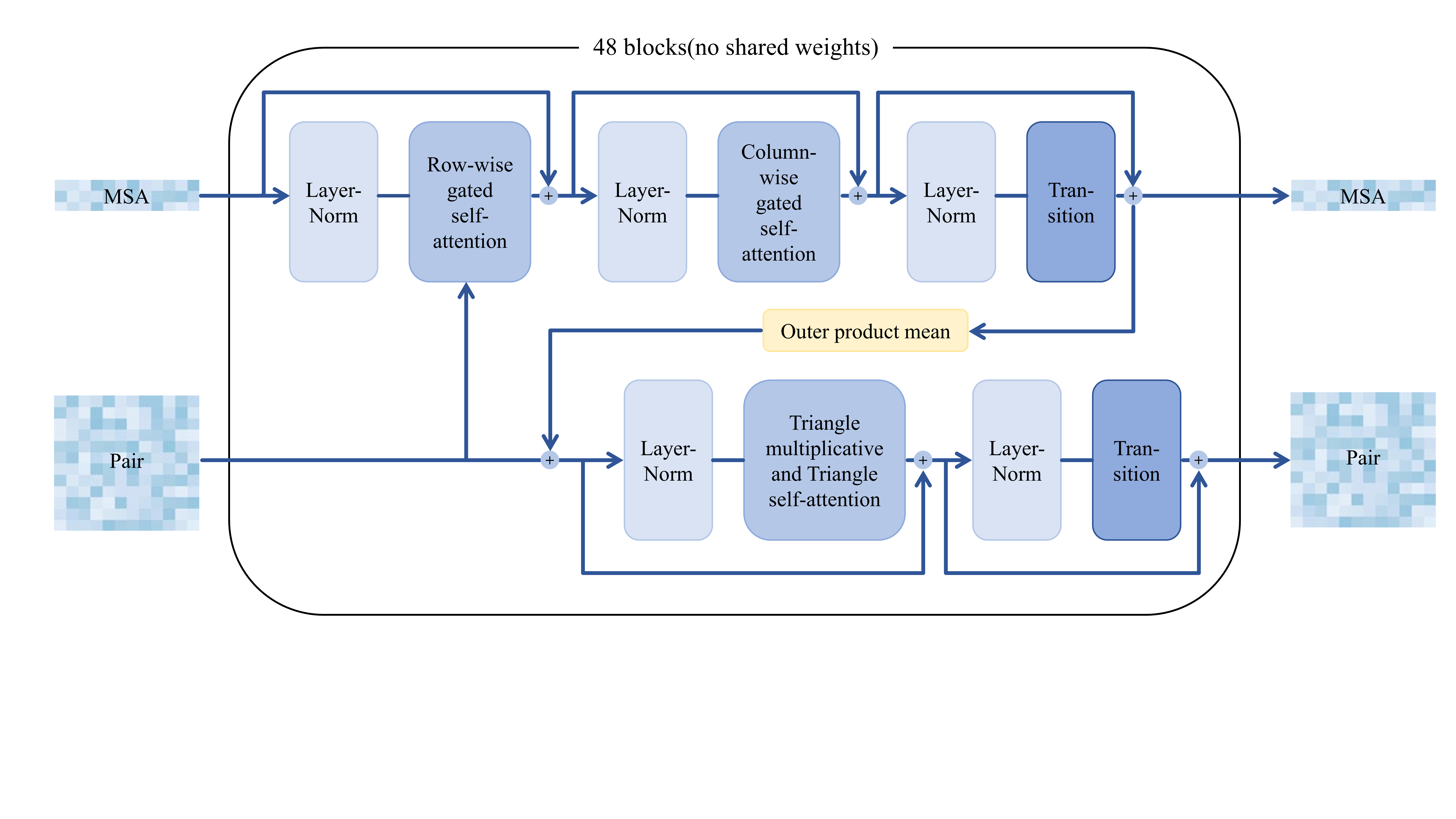}
\caption{The evoformer module of AlphaFold2, which has now been replaced by the pairformer module in AlphaFold3.}
\label{fig6}
\end{figure}

While protein language models excel at capturing evolutionary and structural information, they lack explicit supervision from protein structural data \cite{188prottrans}. To address this, some models integrate structural information in different ways: SaProt \cite{69saprot} converts structural data into 3Di tokens, while ESM-3 \cite{48esm3} fuses sequence, structure, and function modalities into a single latent space. Other models process structure as continuous vectors, such as LM-GVP \cite{10lmgvp}, which connects sequence data to graph-based features, and PeTriBERT \cite{72petribert}, which encodes 3D structural features using Fourier embeddings. MSA information is integrated into models like MSA-Transformer \cite{91msatransformer}, which extends MLM to MSA, and MSA-Augmenter \cite{92msaaugmenter}, which generates homologous sequences. The AlphaFold 2’s Evoformer submodule (Fig.\ref{fig6}), while computationally intensive, is a powerful tool for handling MSA data.

\subsubsection{Decoder-only models}

Decoder-only models, such as GPT, are commonly used for protein generation tasks. ProGen\cite{76progen} is a self-regressive model trained through unsupervised sequence generation tasks, functioning as a controllable protein generation language model. ProGen2\cite{77progen2} first expands the number of parameters to 6.4 billion. RITA\cite{78rita}, trained on UniProt100 data, aiming to accelerate protein design through large-scale model training. ProtGPT2\cite{79protgpt2} is another autoregressive model capable of generating sequences within seconds. 

PoET\cite{80poet} is an autoregressive generative model for protein family distribution, which generates a set of sequences rather than a single sequence to produce specific family proteins. lgLM \cite{81iglm} generates synthetic antibody libraries by redesigning variable-length spans in antibody sequences. ZymCTRL\cite{82zymctrl} is a conditional language model for controllably generating artificial enzymes in zero-shot settings. ProLLaMA\cite{84prollama} introduces a two-stage training framework and becomes the first known model to simultaneously address multiple Protein Language Processing (PLP) tasks with a ProLLM (Protein Large Language Model).

\subsubsection{Encoder-Decoder models}

The encoder-decoder architecture is typically used for sequence-to-sequence tasks. pAbT5\cite{87pabt5} is a T5-based\cite{85t5} model that incorporates biological constraints on chain pairing preferences to generate complementary heavy or light chains from their paired partners, while IgT5 \cite{IgBert} was the first T5 antibody language model trained for sequence encoding to date. xTrimoPGLM\cite{88xtrimopglm} is based on the GLM architecture\cite{86glm}, boasting an unprecedented scale of 100 billion parameters and 1 trillion training tokens, exploring the compatibility and joint optimization potential of simultaneously addressing protein understanding and generation tasks. SS-pLM\cite{89ssplm} introduces a smaller language model and demonstrates that appropriate fine-tuning can achieve performance comparable to larger models. Ankh\cite{90ankh} proposes an auto-regressive fine-tuning generation framework for the High-N (protein family-based generation).

\begin{table*}[p] 
\centering

\begin{tabular}{cccccccccc} 
\toprule
\multicolumn{3}{c}{Model} & Pretraining Dataset & Base Model & Params & Time & Code \\  

\hline

\multicolumn{3}{c}{CARP} & UniRef50 & CNN & 600K-640M & 2024.02 & \href{https://github.com/microsoft/protein-sequence-models}{\ding{51}} \\

\multicolumn{3}{c}{MIF-ST} & CATH & GNN & 3.4M & 2023.03 & 
\href{https://github.com/microsoft/protein-sequence-models}{\ding{51}} \\

\multicolumn{3}{c}{ProSE} & UniRef90, SCOP & LSTM & - & 2021.06 & \href{https://github.com/tbepler/prose}{\ding{51}} \\

\multicolumn{3}{c}{Seq2vec} & - & CNN-LSTM & - & 2020.09 & \ding{55} \\

\multicolumn{3}{c}{UDSMProt} & UniProtKB/Swiss-Prot & AWD-LSTM & - & 2020.01 & \ding{55} \\

\multicolumn{3}{c}{SeqVec} & UniRef50 & ELMo & - & 2019.12 & \href{https://github.com/mheinzinger/SeqVec}{\ding{51}} \\

\multicolumn{3}{c}{UniRep} & UniRef50 & mLSTM & - & 2019.10 & \href{https://github.com/churchlab/UniRep}{\ding{51}} \\

\multicolumn{3}{c}{ProtVecX} & UniRef50, UniProtKB/Swiss-Prot & ProVec & - & 2019.03 & \ding{55} \\

\multicolumn{3}{c}{ProtVec} & UniProtKB/Swiss-Prot & Skip-gram & - & 2015.11 & \ding{55} \\

\bottomrule
\end{tabular}
\caption{Non-transformer-based models}
\label{Non-transformer-based models}

\hspace*{0cm}
\begin{tabular}{cccccccccc} 
\toprule
\multicolumn{3}{c}{Model} & Pretraining Dataset & Params & Time & Code \\  

\hline

\multicolumn{3}{c}{ESM-C} & UniRef, MGnify, JGI & 300M, 600M, 6B & 2024.12 & \ding{55} \\

\multicolumn{3}{c}{IgBert} & OAS & 420M & 2024.12 & \ding{55} \\

\multicolumn{3}{c}{AMPLIFY} & UniRef50, UniRef100, OAS, SCOP & 120M/350M & 2024.09 & \href{https://github.com/chandar-lab/AMPLIFY}{\ding{51}} \\

\multicolumn{3}{c}{ESM-3} & UniRef, MGnify, AlphaFoldDB, ESMAtlas & 98B & 2024.07 & \href{https://github.com/evolutionaryscale/esm}{\ding{51}} \\

\multicolumn{3}{c}{BALM} & OAS & - & 2024.05 & \href{https://github.com/BEAM-Labs/BALM}{\ding{51}} \\

\multicolumn{3}{c}{ESM All-Atom} & AlphaFoldDB & 35M & 2024.05 & \href{https://github.com/zhengkangjie/ESM-AA}{\ding{51}} \\

\multicolumn{3}{c}{AbLang2} & OAS & - & 2024.02 & \href{https://github.com/oxpig/AbLang2.git}{\ding{51}} \\

\multicolumn{3}{c}{ESM-GearNet} & AlphaFoldDB & - & 2023.10 & \href{https://github.com/DeepGraphLearning/ESM-GearNet}{\ding{51}} \\

\multicolumn{3}{c}{ProtFlash} & UniRef50 & 79M/174M & 2023.10 & \href{https://github.com/ISYSLAB-HUST/ProtFlash}{\ding{51}} \\

\multicolumn{3}{c}{SaProt} & AlphaFoldDB, PDB & 650M & 2023.10 & \href{https://github.com/westlake-repl/SaProt}{\ding{51}} \\

\multicolumn{3}{c}{ESM-2} & UniRef50 & 8M-15B & 2023.03 & \href{https://github.com/facebookresearch/esm}{\ding{51}} \\

\multicolumn{3}{c}{PromptProtein} & UniRef50, PDB & 650M & 2023.02 & \href{https://github.com/HICAI-ZJU/PromptProtein}{\ding{51}} \\

\multicolumn{3}{c}{AminoBert} & UniRef90, PDB, MGnify & - & 2022.10 & \ding{55} \\

\multicolumn{3}{c}{DistilProtBert} & UniRef50 & 230M & 2022.09 & \href{https://github.com/yarongef/DistilProtBert}{\ding{51}} \\

\multicolumn{3}{c}{PeTriBERT} & AlphaFoldDB & 40M & 2022.08 & \ding{55} \\

\multicolumn{3}{c}{AntiBERTa} & OAS & 86M & 2022.07 & \href{https://github.com/alchemab/antiberta}{\ding{51}} \\

\multicolumn{3}{c}{AbLang} & OAS & - & 2022.06 & \href{https://github.com/oxpig/AbLang}{\ding{51}} \\

\multicolumn{3}{c}{OntoProtein} & ProteinKG25 & - & 2022.06 & \href{https://github.com/zjunlp/OntoProtein}{\ding{51}} \\

\multicolumn{3}{c}{LM-GVP} & - & - & 2022.04 & \href{https://github.com/aws-samples/lm-gvp}{\ding{51}} \\

\multicolumn{3}{c}{ProteinBERT} & UniRef90 & 16M & 2022.03 & \href{https://github.com/nadavbra/protein_bert}{\ding{51}} \\

\multicolumn{3}{c}{AntiBERTy} & OAS & 26M & 2021.12 & \href{https://pypi.org/project/antiberty/}{\ding{51}} \\

\multicolumn{3}{c}{ProteinLM} & Pfam & 200M/3B & 2021.12 & \href{https://github.com/THUDM/ProteinLM}{\ding{51}} \\

\multicolumn{3}{c}{TCR-BERT} & PIRD, VDJdb, TCRdb, murine LCMV GP33 & 100M & 2021.11 & \href{https://github.com/wukevin/tcr-bert}{\ding{51}} \\

\multicolumn{3}{c}{PMLM} & UniRef50, Pfam & 87M-731M & 2021.10 & \ding{55} \\

\multicolumn{3}{c}{ProtTrans} & UniRef, BFD & - & 2021.07 & \href{https://github.com/agemagician/ProtTrans}{\ding{51}} \\

\multicolumn{3}{c}{ESM-MSA-1b} & UniRef50 & 100M & 2021.02 & \href{https://github.com/facebookresearch/esm}{\ding{51}} \\

\multicolumn{3}{c}{ESM-1v} & UniRef90 & 650M & 2021.02 & \href{https://github.com/facebookresearch/esm}{\ding{51}} \\

\multicolumn{3}{c}{PRoBERTa} & UniProtKB/Swiss-Prot & 44M & 2020.09 & \href{https://github.com/annambiar/PRoBERTa}{\ding{51}} \\

\multicolumn{3}{c}{ESM-1b} & UniRef50 & 650M & 2020.02 & \href{https://github.com/facebookresearch/esm}{\ding{51}} \\

\bottomrule
\end{tabular}
\caption{Encoder-only models}
\label{Encoder-only models}

\begin{tabular}{cccccccccc} 
\toprule
\multicolumn{3}{c}{Model} & Pretraining Dataset & Params & Time & Code \\  

\hline

\multicolumn{3}{c}{ProLLaMA} & UniRef50 & - & 2024.02 & \href{https://github.com/PKU-YuanGroup/ProLLaMA}{\ding{51}} \\

\multicolumn{3}{c}{PoET} & - & 57M-604M & 2023.11 & \ding{55} \\

\multicolumn{3}{c}{ProGen2} & UniRef90, BFD30, PDB & 151M-6.4B & 2023.10 & \href{https://github.com/salesforce/progen}{\ding{51}} \\

\multicolumn{3}{c}{IgLM} & - & 13M & 2022.12 & \href{https://github.com/Graylab/IgLM}{\ding{51}} \\

\multicolumn{3}{c}{RITA} & UniRef100 & 1.2B & 2022.05 & \ding{55} \\

\multicolumn{3}{c}{DARK} & - & 128M & 2022.01 & \ding{55} \\

\multicolumn{3}{c}{ZymCTRL} & BRENDA & 738M & 2022.01 & \href{https://huggingface.co/AI4PD/ZymCTRL}{\ding{51}} \\

\multicolumn{3}{c}{ProtGPT2} & UniRef50 & 738M & 2021.01 & \href{https://huggingface.co/nferruz/ProtGPT2}{\ding{51}} \\

\multicolumn{3}{c}{ProGen} & UniParc, UniProtKB/Swiss-Prot & 1.2B & 2020.03 & \href{https://github.com/salesforce/progen}{\ding{51}} \\

\bottomrule
\end{tabular}
\caption{Decoder-only models}
\label{Decoder-only models}

\begin{tabular}{cccccccccc} 
\toprule
\multicolumn{3}{c}{Model} & Pretraining Dataset & Params & Time & Code \\  

\hline

\multicolumn{3}{c}{IgT5} & OAS & 3B & 2024.12 & \ding{55} \\

\multicolumn{3}{c}{ProSST} & AlphaFoldDB, CATH & 110M & 2024.05 & \href{https://github.com/ai4protein/ProSST}{\ding{51}} \\

\multicolumn{3}{c}{pAbT5} & OAS & - & 2023.10 & \ding{55} \\

\multicolumn{3}{c}{SS-pLM} & UniRef50 & 14.8M & 2023.08 & \ding{55} \\

\multicolumn{3}{c}{ProstT5} & AlphaFoldDB, PDB & 3B & 2023.07 & \href{https://github.com/mheinzinger/ProstT5}{\ding{51}} \\

\multicolumn{3}{c}{xTrimoPGLM} & UniRef90, ColdFoldDB & 100B & 2023.07 & \ding{55} \\

\multicolumn{3}{c}{MSA-Augmenter} & UniRef50 & 260M & 2023.06 & \href{https://github.com/lezhang7/MSA-Augmentor}{\ding{51}} \\

\multicolumn{3}{c}{LM-Design} & - & 664M & 2023.02 & \ding{55} \\

\multicolumn{3}{c}{Ankh} & UniRef50 & 450M/1.15B & 2023.01 & \href{https://github.com/agemagician/Ankh}{\ding{51}} \\

\multicolumn{3}{c}{ProtT5} & UniRef50, BFD & 3B/11B & 2022.06 & \href{https://huggingface.co/Rostlab/prot_t5_xl_uniref50}{\ding{51}} \\

\multicolumn{3}{c}{Sapiens} & OAS & 0.6M & 2022.02 & \href{https://github.com/Merck/BioPhi}{\ding{51}} \\

\bottomrule
\end{tabular}
\caption{Encoder-decoder models}
\label{Encoder-decoder models}
\end{table*}

To combine sequence and structure information, a common way to handle structural information is to encode it into discrete tokens. ProstT5\cite{93prostt5} proposes a bilingual translation architecture that allows mutual conversion between 3Di tokens and sequence embedding tokens. ProSST\cite{95prosst} utilizes a structure quantization module to obtain structural tokens, employing a disentangled attention mechanism to assign different weights to structure and sequence tokens. Other models process structure as continuous vectors, such as LM-Design\cite{96lmdesign}, which uses a structural encoder to process structural information independently of sequence information, and then introduces a lightweight structural adapter to integrate structural and sequence information.

\section{Position Encoding}
\label{Position Encoding}

Transformer does not explicitly model relative or absolute positional information within its structure, necessitating the introduction of explicit position encodings\cite{1transformer}. Given the widespread application of Transformer in protein language models (PLMs), it is important to discuss commonly used position encodings. In this section, we introduce the two main types of positional encoding used in protein language modeling, including absolute positional encoding, relative positional encoding, and their variants.

\subsection{Absolute Position Encoding}

Absolute position encoding is one of the earliest forms of position encoding, allowing for explicit modeling of absolute positional information by adding absolute position encodings to the input\cite{97}. There are many choices of positional encodings, learned and fixed\cite{1transformer,39bert}. However, Learned positional encoding often leads to better downstream performance for protein language models, as adopted by numerous PLMs like ESM-1b\cite{5esm1b}, and others\cite{53distilprotbert,79protgpt2,91msatransformer}. Absolute position encoding is widely used due to its simplicity and computational efficiency, yet it requires encoding for each position, limiting the model’s ability to handle sequences of varying lengths and lacking extrapolation capability\cite{97,42protflash}. Additionally, Sinha et al. (2022)\cite{105} explored the limitations of absolute position encoding, noting that Transformer models may overly rely on positional information in certain cases, leading to significant performance drops when the positions of input sentences shift.

Rotary position encoding (RoPE)\cite{107} encodes absolute positions using rotation matrices while incorporating explicit relative position dependencies into the self-attention formula, achieving a unification of the two. RoPE provides valuable features, including flexibility in sequence length, attenuated dependencies between tokens as relative distances increase, and the capability to equip linear self-attention with relative position encoding\cite{107}. These advantages have led to the use of rotary position encoding in numerous prominent PLMs, such as ProtFlash\cite{42protflash}, ESM-2\cite{47esm2}, ProGen2\cite{77progen2}, and RITA\cite{78rita}. Among these, RITA\cite{78rita} compared RoPE and ALIBI position encodings, ultimately determining that RoPE yielded superior results in language modeling loss. ESM-2\cite{47esm2} observed that RoPE improved the quality of smaller models, although performance enhancements began to diminish with increased model size and training time. 

\subsection{Relative Position Encoding}

Relative position embeddings do not use fixed embeddings for each position; instead, they generate different learned embeddings based on the offsets between ‘keys’ and ‘queries’ in the self-attention mechanism\cite{85t5}. Compared to absolute position encoding, relative position encoding is insensitive to the total length of the sequence, making it better suited for handling sequences of arbitrary lengths and capturing structural information\cite{47esm2,97}. Several variants\cite{100transformerxl,85t5,102deberta,103alibi} have been born based on the classical relative position coding of Shaw et al.\cite{97}. Although relative position encoding has become increasingly common in recent years\cite{85t5}, some studies indicate that absolute position encoding remains crucial in certain scenarios\cite{106,102deberta}, as the widespread use of hybrid positional coding of relative positional coding and rotational positional coding in protein language modeling\cite{85t5,107}. DeBERTa\cite{102deberta} indicates that the position at which absolute position encoding is introduced when using mixed encoding affects the learning of relative position encoding.

\section{Scaling Laws}
\label{Scaling Laws}
In recent years, there has been a growing trend in machine learning to scale neural networks, a phenomenon particularly evident in the field of PLMs\cite{108}. Scaling up PLMs has proven to be significantly more beneficial than scaling up language models in natural language processing \cite{78rita}. Scaling Laws\cite{109}, a theory proposed and validated by OpenAI during the development of the Generative Pre-trained Transformer series\cite{110gpt1,111gpt2,112gpt3}, describe how performance is influenced by model size, dataset size, and the amount of compute used for training. The ESM series\cite{5esm1b,47esm2,48esm3} clearly demonstrates the performance improvements associated with increasing model size, suggesting that the principles of the scaling law also apply to protein language modeling. Moreover, several studies\cite{76progen,78rita,5esm1b} have observed that the modeling loss of PLMs typically follows a strikingly close power law relationship, and that PLMs are more prone to underfitting. Even when PLMs are trained far beyond the optimal point in NLP, they still appear undertrained, indicating the need for larger-scale training. These findings suggest that further scaling of models can substantially improve the performance of PLMs.

\section{Datasets}
\label{Datasets}

In this section, we will review commonly used pretraining datasets and benchmark datasets, which are used respectively for self-supervised training and model evaluation, and classify them based on their applicable scopes. Finally, we summarize the datasets and benchmarks in Tables\ref{Datasets} and Table\ref{Benchmarks}.

\subsection{Pre-training Datasets}

\subsubsection{Sequence Datasets}

\subsubsection*{\bf UniProt dataset}

The UniProt Knowledgebase\cite{116uniprot,120uniprot} aims to provide users with a comprehensive, high-quality, and freely accessible set of protein sequences annotated with functional information. It comprises multiple sub-databases, each focusing on different types of information.

\begin{itemize}
\item{UniProtKB}
\end{itemize}

UniProtKB\cite{117uniprotkb,14uniprotkbswissprot} consists of two components, Swiss-Prot and TrEMBL, which provide high-quality manual annotation and automatic publication of protein sequences, respectively.

\begin{itemize}
\item{UniRef}
\end{itemize}

UniRef(UniRef100, UniRef90, UniRef50)\cite{26uniref,27uniref} is one of the most widely used protein sequence databases, constructed as a clustering system based on the UniProt protein database. Sequences in UniRef100 are clustered at 100\% similarity, while UniRef90 and UniRef50 are created by clustering sequences at 90\% and 50\% similarity thresholds, respectively. UniRef50 and UniRef100 are widely used for training protein language models (PLMs)\cite{5esm1b,32seqvec,21carp,91msatransformer,188prottrans,78rita,108} due to low redundancy and comprehensive sequence distribution, respectively. UniRef90 is widely used in mutation prediction models like ESM-1v\cite{7esm1v} for capturing subtle differences caused by mutations. Ankh\cite{90ankh} and AMPLIFY\cite{108} compared the differences in the use of these three different similarity-based clustering databases for model training.

\begin{itemize}
\item{Uniclust}
\end{itemize}

The Uniclust databases(Uniclust90, Uniclust50, Uniclust30)\cite{122uniclust} cluster UniProtKB sequences at levels of 90\%, 50\%, and 30\% pairwise sequence identity using MMseq2. Uniclust90 and Uniclust50 clusters demonstrated better consistency in functional annotation compared to UniRef90 and UniRef50. Meanwhile, Uniclust30, due to its low similarity, is widely used for homologous sequence-related tasks\cite{79protgpt2,128bfd}

\begin{itemize}
\item{UniParc(UniProt Archive)}
\end{itemize}

UniParc\cite{123uniparc,116uniprot} aims to offer a complete set of known unique sequences, rather than detailed functional annotation information like other UniProt datasets.

\subsubsection*{\bf Pfam}

Pfam\cite{125pfam,126pfam} is a widely used protein family and domain database, with each family defined by two alignments and a profile Hidden Markov Model (HMM). Pfam is extensively utilized in research regarding function, structure, and evolution. Models like ProteinLM\cite{54proteinlm}, TAPE\cite{99tape}, and ESM-1b\cite{5esm1b} have been trained on Pfam.

\subsubsection*{\bf BFD(Big Fantastic Database)}

BFD\cite{49bfd,128bfd} is a massive database containing hundreds of millions of protein sequences, integrating data from various public databases. It has been used for training multiple PLMs\cite{77progen2,188prottrans}, providing MSA information for models like AlphaFold2\cite{128bfd}.

\subsubsection*{\bf MGnify}

MGnify database\cite{129mgnify} contains 2.4 billion non-redundant protein sequences predicted from metagenomic datasets. These machine-predicted protein sequences are widely employed in PLMs like ESM-2 and ESM-3 for exploring new protein information\cite{47esm2} and enhance the diversity of training data\cite{48esm3}.

\subsubsection*{\bf BRENDA}

Established in 1987, the BRENDA enzyme database\cite{132brenda} has developed into a major enzyme function database system, classifying enzyme information according to the Enzyme Commission (EC) nomenclature. The BRENDA database is mainly used by enzyme-related models like ZymCTRL\cite{82zymctrl} for training.

\subsubsection*{\bf OAS}

OAS\cite{OAS} collected Ig-seq outputs from 55 studies, covering more than half a billion Ab sequences across diverse immune states, organisms (primarily human and mouse), and individuals. These sequences were sorted, cleaned, annotated, translated, and numbered.

\subsubsection*{\bf proteinKG25}

ProteinKG25\cite{ontoprotein} is a novel large-scale knowledge graph(KG) dataset, which contains about 612,483 entities, 4,990,097 triples, and aligned node descriptions from GO annotations, making it the first large-scale KG dataset to facilitate protein pre-training.

\subsubsection*{\bf PIRD}

The Pan Immune Repertoire Database (PIRD)\cite{PIRD} is developed to collect and store annotated TCR and BCR sequencing data, including from Homo sapiens and other species, with a manually curated database of TCRs and BCRs targeting known antigens (TBAdb) also deposited.

\subsubsection*{\bf TCRdb}
TCRdb\cite{TCRdb} is a comprehensive human TCR sequences database, which contains more than 277 million highly reliable TCR sequences from over 8265 TCR-Seq samples across hundreds of tissues/clinicalconditions/cell types, by a uniform pipeline to characterize TCR sequences on TCR-Seq data.

\subsubsection{Structure Datasets}

\subsubsection*{\bf PDB(Protein Data Bank)}

PDB\cite{133pdb} is a widely used database of biomolecular structures, which are obtained experimentally, ensuring high reliability and accuracy. These structures are determined using experimental methods like X-ray crystallography,
 NMRspectroscopy, and, more recently, cryo-electron microscopy.

\subsubsection*{\bf AlphaFoldDB}

Developed and maintained by DeepMind, AlphaFoldDB\cite{134alphafolddb} is a database specifically designed to store and provide protein structures predicted by the AlphaFold series. Although these structures are not experimentally confirmed, the outstanding performance of AlphaFold in structural prediction lends high confidence to these predictions; therefore, it is widely used in structure-related models\cite{72petribert,48esm3,69saprot,93prostt5} as a supplement to the scarce experimental structural.

\subsubsection*{\bf ESMAtlas}

Developed by Meta AI, ESMAtlas\cite{ESMAtlas} is a metagenomic protein structure prediction database powered by ESMFold. The database contains over 617 million predicted protein structures, including approximately 225 million high-confidence results, among which millions of predicted structures are novel compared to experimentally determined ones. This provides new perspectives for the study of previously unknown protein structures.

\subsection{Benchmarks}

\subsubsection{Structure Prediction}

\subsubsection*{\bf CASP(Critical Assessment of Techniques for Protein Structure Prediction)}

CASP\cite{135casp} is essentially a biennial competition aimed at advancing methods for predicting three-dimensional protein structures from amino acid sequences, which provides a unified benchmark for evaluating the performance of different methods in protein structure prediction.

\subsubsection*{\bf CAMEO(Continuous Automated Model EvaluatiOn)}

CAMEO platform\cite{137cameo,138cameo} complements the biennial CASP experiment by conducting fully automated blind evaluations of three-dimensional protein prediction servers based on the weekly prerelease of sequences of those structures. CAMEO is a useful tool for PLMs\cite{47esm2,91msatransformer,163} in structure prediction and contact prediction tasks.

\subsubsection*{\bf SCOP(Structural Classification of Proteins)}

SCOP\cite{139scop,140scop,142scopcath} is a protein structure classification database. It organizes the domains of known protein structures based on structural and evolutionary relationships in a hierarchical manner. SCOP is divided into four levels: class, fold, superfamily, and family. SCOP places greater emphasis on the evolutionary relationships of protein structures and is primarily based on manual classification. ProstT5\cite{93prostt5} detected remote homology in SCOPe40.

\subsubsection*{\bf CATH(Class, Architecture, Topology, Homologous superfamily)}

CATH\cite{141cath,142scopcath} is a protein structure classification database that systematically classifies proteins based on their tertiary structure. The four letters in the database's name represent the four main levels of the classification system: class, architecture, topology, and homologous superfamily. CATH emphasizes the topological classification of protein structures, combining automation with manual curation. ProstT5\cite{93prostt5} useed CATH for structure classification tasks.

\subsubsection{Function}

\subsubsection*{\bf EC(Enzyme Commission Numbers)}

EC dataset\cite{83ec} is a standardized dataset for describing enzyme classification and function, managed and updated by the ExplorEnz database. The EC database contains a wealth of enzyme data, with each enzyme corresponding to an EC number. The EC number is a numerical classification scheme for enzymes, specifying their catalytic functions with a four-digit code, making it an indispensable tool for PLMs\cite{144esmgearnet,69saprot,145clean} in enzyme function prediction.

\subsubsection*{\bf GO(Gene Ontology)}

The GO dataset\cite{146go,147go} is a system aimed at providing standardized descriptions of gene and protein functions. The GO ontology encompasses three main domains: biological process (BP), molecular function (MF), and cellular component (CC). GO annotations cover multiple species and play a positive role in cross-species gene function and evolutionary research, making it a useful tool for PLMs \cite{10lmgvp,58proteinbert}in protein function prediction.

\subsubsection*{\bf FLIP}

FLIP\cite{148flip} is a benchmark for function prediction, designed to encourage rapid scoring of representation learning for protein engineering. Unlike CAFA, FLIP focuses on relevant metrics for protein engineering, including experimental data on the stability of adeno-associated viruses for gene therapy, the stability of protein domain B1, immunoglobulin binding, and thermal stability across multiple protein families\cite{69saprot}.

\subsubsection*{\bf CAFA}

CAFA\cite{150cafa,151cafa} is a large scientific competition evaluating the accuracy of bioinformatics tools and methods, aimed at evaluating and improve computational annotation methods for protein functions by comparing the gene ontology (GO) functional terms predicted by computational methods with subsequent experimental validation annotations. CAFA serves as an important benchmark for function prediction, widely used in PLMs related to function prediction\cite{152atgo,153sprofgo,154deepfri,155phignet}.

\subsubsection{Others}

\subsubsection*{\bf TAPE(Tasks Assessing Protein Embeddings)}

TAPE\cite{99tape} offers a series of carefully designed tasks and datasets, covering five biologically relevant semi-supervised learning tasks including secondary structure prediction, contact prediction, remote homology prediction, fluorescence prediction, and stability prediction, aimed at evaluating protein embeddings\cite{69saprot,60pmlm}. 

\subsubsection*{\bf PEER}

PEER\cite{149peer} is a comprehensive, multi-task benchmark for protein sequence understanding. It provides a variety of different protein understanding tasks, including protein function prediction, protein localization prediction, protein structure prediction, protein-protein interaction prediction, and protein-ligand interaction prediction\cite{69saprot}. 

\subsubsection*{\bf ProteinGym}

ProteinGym\cite{156proteingym} conducts deep mutational scanning (DMS) analyses, covering millions of mutant sequences and providing high-quality clinical datasets. ProteinGym offers two types of mutation analyses: substitution and insertion/deletion mutations, for evaluating machine learning models' performance in predicting protein mutation effects in both zero-shot and supervised environments. ProteinGYM is widely used in PLMs\cite{69saprot,48esm3,104tranception,190proteinnpt} for protein mutation assessments.

\begin{table*}[!htbp] 
\centering
\begin{tabular}{ccccc} 
\toprule

\multicolumn{2}{c}{Pre-training datasets}&Time&Scale&Link\\

\hline

\multirow{11}*{Sequence datasets}
&{BFD}&2021.07&2.5B&\href{https://bfd.mmseqs.com/}{\ding{51}}\\

&{BRENDA}&2002.01&-&\href{https://www.brenda-enzymes.org/}{\ding{51}}\\

&{MGnify}&2022.12&-&\href{https://www.ebi.ac.uk/metagenomics/}{\ding{51}}\\

&{Pfam}&2023.09&47M&\href{https://www.ebi.ac.uk/interpro/entry/pfam/}{\ding{51}}\\

&{Uniclust30}&2016.11&-&\href{https://uniclust.mmseqs.com/}{\ding{51}}\\

&{UniParc}&2023.11&632M&\href{https://www.uniprot.org/uniparc?query=*}{\ding{51}}\\

&{UniProtKB/Swiss-Prot}&2023.11&570K&\href{https://www.uniprot.org/uniprotkb?query=*}{\ding{51}}\\

&{UniProtKB/TrEMBL}&2023.11&251M&\href{https://www.uniprot.org/uniprotkb?query=*}{\ding{51}}\\

&{UniRef50}&2023.11&53M&\href{https://www.uniprot.org/uniref?query=*}{\ding{51}}\\

&{UniRef90}&2023.11&150M&\href{https://www.uniprot.org/uniref?query=*}{\ding{51}}\\

&{UniRef100}&2023.11&314M&\href{https://www.uniprot.org/uniref?query=*}{\ding{51}}\\

\hline

\multirow{2}*{Structural datasets}

&{AlphaFoldDB}&2021.11&200M&\href{https://alphafold.ebi.ac.uk/}{\ding{51}}\\

&{PDB}&2023.12&214K&\href{https://www.rcsb.org/}{\ding{51}}\\

\bottomrule

\end{tabular}
\caption{Pre-training datasets}
\label{Pre-training datasets}
\end{table*}

\begin{table*}[!htbp] 
\centering
\begin{tabular}{ccccc} 
\toprule

\multicolumn{2}{c}{Benchmarks}&Time&Scale&Link\\

\hline

\multirow{4}*{Structural benchmarks}
&{CAMEO}&-&-&\href{https://cameo3d.org/}{\ding{51}}\\

&{CASP}&-&-&\href{https://predictioncenter.org/}{\ding{51}}\\

&{CATH}&2023.02&151M&\href{http://www.cathdb.info/}{\ding{51}}\\

&{SCOP}&2023.01&914K&\href{http://scop.berkeley.edu/}{\ding{51}}\\

\hline

\multirow{4}*{Functional benchmarks}

&{CAFA}&-&-&\href{https://biofunctionprediction.org/}{\ding{51}}\\

&{EC}&2023.11&2.6M&\href{https://www.enzyme-database.org/}{\ding{51}}\\

&{FLIP}&2022.01&320K&\href{https://benchmark.protein.properties/}{\ding{51}}\\

&{GO}&2023.11&1.5M&\href{https://geneontology.org/}{\ding{51}}\\

\hline

\multirow{3}*{Other benchmarks}

&{PEER}&2022.11&390K&\href{https://github.com/DeepGraphLearning/PEER_Benchmark}{\ding{51}}\\

&{ProteinGym}&2022.12&300K&\href{https://proteingym.org/}{\ding{51}}\\

&{TAPE}&2021.09&120K&\href{https://github.com/songlab-cal/tape}{\ding{51}}\\

\bottomrule
\end{tabular}
\caption{Benchmarks}
\label{Benchmarks}
\end{table*}

\section{Applications of Protein Language Models}
\label{Applications of Protein Language Models}

\begin{figure}[!h]
\centering
\includegraphics[scale=0.26 ]{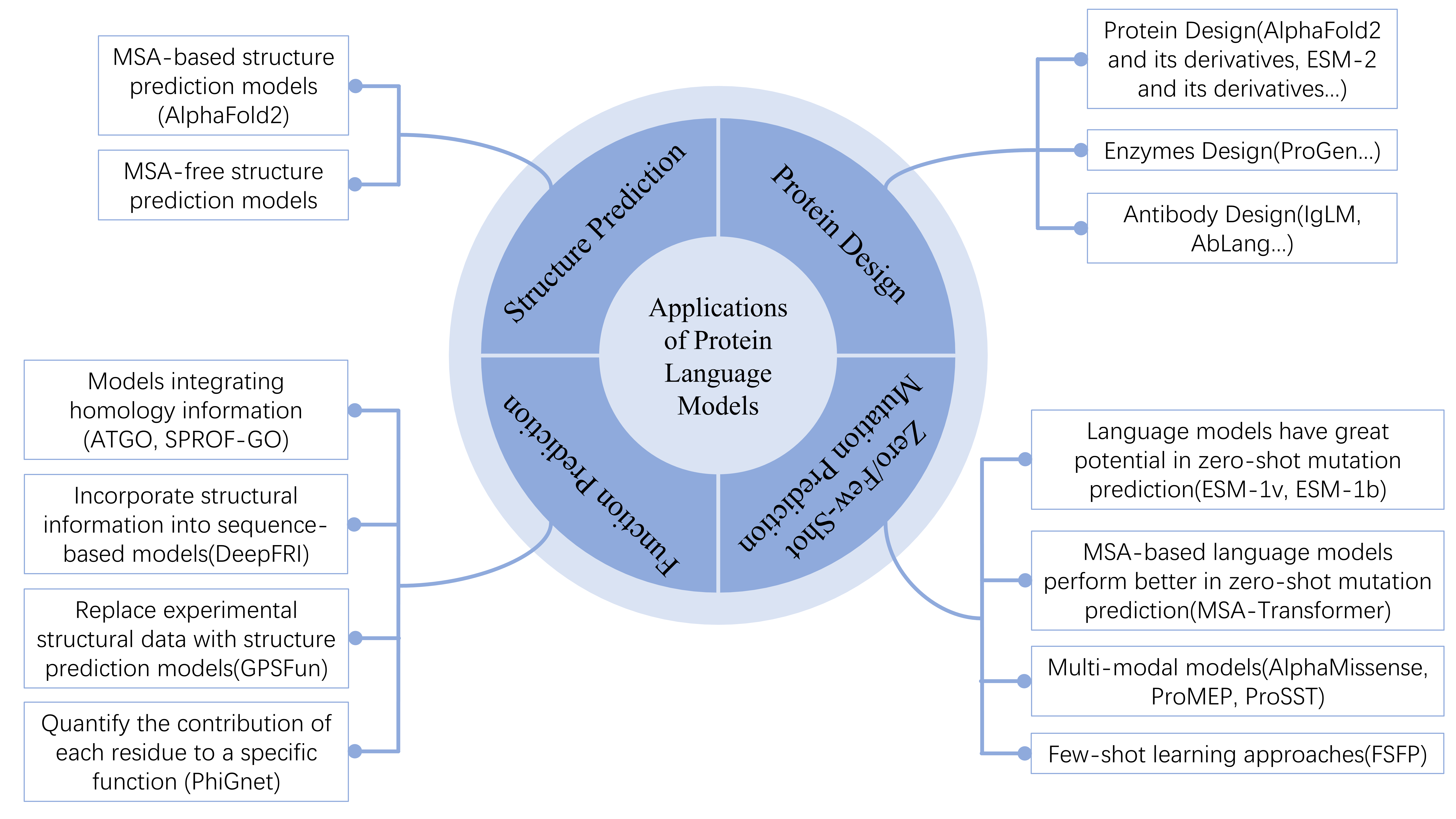}
\caption{Applications of protein language models.}
\label{fig5}
\end{figure}

Protein language models, through deep learning on large datasets of protein sequences, are able to capture the underlying functional and structural information hidden within these sequences\cite{5esm1b}, leading to applications in various downstream tasks (Fig.\ref{fig5}), including structure prediction, function prediction, protein design, and mutation prediction.

\subsection{Structure Prediction}

Predicting the three-dimensional (3D) structure of proteins from amino acid sequences has traditionally relied on physics-based approaches\cite{59aminobert}. About a decade ago, the research focus shifted to extracting residue-residue contacts from co-evolutionary relationships embedded in multiple sequence alignments (MSA)\cite{158}. A representative model leveraging MSA, AlphaFold2\cite{128bfd}, achieved groundbreaking success in the CASP14 competition, demonstrating that MSA can substantially enhance the performance of protein structure prediction models. However, the performance of MSA-based models deteriorates significantly in the absence of homologous protein sequences. Although various strategies have been proposed to address the limitations posed by the lack of MSA\cite{92msaaugmenter,162}, a more fundamental solution involves reducing or even eliminating the reliance on MSA altogether. Indeed, the mainstream direction in recent developments has been to create MSA-independent structural prediction models\cite{161}, with an increasing number of such models being reported\cite{59aminobert,47esm2,165,166,163,164}. These MSA-free models have shown strong performance on orphan proteins\cite{59aminobert,166} and significant advantages over single-sequence versions of AlphaFold2 and RoseTTAFold\cite{47esm2,164,165}. Recent studies\cite{47esm2,163} have demonstrated that MSA-free models can be trained on large-scale datasets to implicitly learn co-evolutionary information, allowing them to bypass MSA. Furthermore, MSA-free models exhibit substantially faster computational speeds, addressing the bottleneck of slow MSA retrieval\cite{159}. Based on these advancements, we hypothesize that algorithms capable of folding proteins from a single sequence will provide novel insights into protein biophysics.

\subsection{Function Prediction}

Early methods for predicting protein function were heavily constrained by homology relationships, which led to the exploration of utilizing the rich embedding information provided by protein language models as a novel alternative\cite{167}. ATGO\cite{152atgo} and SPROF-GO\cite{153sprofgo} further integrate homology information to enhance prediction accuracy. Given that protein function depends on their unique three-dimensional structures, sequence-based models like DeepFRI\cite{154deepfri} have attempted to incorporate structural information. GPSFun\cite{168gpsfun} demonstrated the feasibility of replacing experimental structural data with structure prediction models like ESMFold, addressing the issue of the scarcity of experimental protein data. Since most models focus solely on the impact of sequence on function, PhiGnet\cite{155phignet} introduced a method to quantify the contribution of each residue to a specific function, offering promising new insights into the interpretability of function prediction.

\subsection{Protein Design}

De novo protein design, a central goal of synthetic biology, has long been hindered by the challenge of reliably predicting the 3D structure of proteins from their amino acid sequences\cite{pd1,pd2,pd3}. However, the advent of protein language models has shown great promise in overcoming this barrier. AlphaFold2 and its derivatives have demonstrated significant potential in protein design, particularly in structure prediction\cite{pd6,pd7,pd8,pd9,pd10,pd11,pd12,pd13,pd14}. Meanwhile, the development of models like the ESM series has broadened the scope of protein design by providing novel tools for sequence-structure-function relationships\cite{pd15,pd16,pd17,pd18,48esm3,pd10,pd21,pd22,pd23}. The ProGen model, for example, generates functional lysozymes with catalytic efficiencies close to those of natural enzymes, despite low sequence identity to natural proteins\cite{pd24}. The IgLM model, on the other hand, optimizes antibody sequence design by leveraging bidirectional contextual information and large antibody sequence datasets\cite{81iglm}. Similarly, the PeTriBERT model incorporates 3D structural representations to design novel proteins with green fluorescent protein (GFP)-like structures\cite{pd26}. Moreover, the Sapiens method, which specializes in antibody humanization, has shown performance comparable to human experts\cite{pd27}. The AbLang model focuses on antibody sequence refinement\cite{pd28}, while AbLang2 \cite{AbLang2} was designed to overcome the effects of the germline bias found in the OAS dataset. The ReLSO autoencoder jointly generates sequences and predicts adaptiveness, allowing for the generation of more functional protein sequences\cite{pd29}. DARK, another notable model, exhibits strong capabilities in complex protein structure generation tasks, demonstrating the power of deep learning in this domain\cite{pd30}. ProtGPT2, built on the GPT-2 architecture, explores new sequence spaces by generating novel protein sequences\cite{79protgpt2}. Furthermore, Biswas et al. combined supervised language models with Markov Chain Monte Carlo methods to optimize sequences for green fluorescent protein and $\beta$-lactamase, leading to significant improvements in their functionality\cite{pd32}. In addition to these, while RoseTTAFold is not strictly a protein language model, its applications in protein design, largely due to the pioneering work of David Baker, have made it an important tool in this field\cite{pd33,pd34,pd35}.

\subsection{Zero/Few-Shot Mutation Prediction}

Traditional methods for mutation prediction require training a new model for each prediction task, which is highly inconvenient. As a result, research has shifted toward using protein language models (PLMs) to predict the effects of zero-shot mutations, meaning inferring the impact of mutations on protein fitness without experimental data support\cite{7esm1v}. Liu et al.\cite{171deepsequence} found that PLMs (particularly ESM-1v and ESM-1b) outperformed the non-language model EVE\cite{172eve} in predicting the pathogenicity of cancer-related mutations and patient survival rates. Joshua's findings showed that ESM-1v performed comparably to the non-language model DeepSequence on deep mutation datasets\cite{7esm1v,171deepsequence}. These results demonstrate the significant potential of language models in zero-shot mutation prediction. Some MSA-based PLMs have even performed better: in zero-shot mutation scenarios, MSA-Transformer outperformed strong models like ESM-1v\cite{171deepsequence}, indicating that incorporating MSA can enhance the zero-shot prediction capabilities of language models. Multi-modal models such as AlphaMissense\cite{173alphamissense}, ProMEP\cite{174promep}, and ProSST\cite{95prosst} have achieved state-of-the-art performance by integrating sequence and structural information. Additionally, few-shot learning approaches (FSFP)\cite{38} can leverage small amounts of experimental data for targeted fine-tuning, significantly improving the prediction accuracy of PLMs.

\section{Tools}
\label{Tools}

In the field of protein sequence analysis and structure prediction, a multitude of efficient tools and methods have emerged, significantly advancing the frontier of protein research. These tools encompass various aspects, including sequence alignment, structural comparison, clustering analysis, and visualization.

\subsubsection{Sequence}

\subsubsection*{\bf MMseq2}

MMseqs2\cite{178mmseqs2} is a set of tools for rapid and efficient protein sequence alignment and clustering, supporting sensitive sequence search and analysis on large-scale datasets. It offers various functionalities, focusing on search, clustering, and database creation\cite{49bfd,122uniclust}.

\subsubsection*{\bf HHblits}

HHblits\cite{179hhblits} is a protein sequence alignment and homology detection tool. It represents query and database sequences using hidden Markov models (HMMs), which serve as a compact representation of multiple sequence alignments (MSAs). 

\subsubsection*{\bf Jackhmmer}

jackhmmer\cite{jackhmmer} is a tool in the HMMER software package used to search for sequence homology in protein databases. It uses sequences, alignments, or profile HMMs for the search. In the first iteration, its search is similar to phmmer; in later iterations, it combines the query’s alignment with target sequences that meet the inclusion threshold, builds a profile, and completes the profile search. Jackhmmer is designed to detect distant homologous sequences with high sensitivity, leveraging its probabilistic models.

\subsubsection{Structure}

\subsubsection*{\bf PyMOL}

PyMOL\cite{180pymol,181pymol} is a cross-platform visualization and analysis tool for biomolecules. It can display the complete 3D structure of proteins, supporting file formats such as PDB and mmCif, and calculating common geometric parameters like RMSD. 

\subsubsection*{\bf TM-align}

TM-align\cite{182tmalign} is a widely used structural alignment tool for comparing and aligning the 3D structures of two proteins. Its core functionality is to calculate the optimal alignment by maximizing the structural similarity between the two protein structures, generating a score called the TM-score (Template Modeling Score) to measure their similarity. 

\subsubsection*{\bf Foldseek}

Foldseek\cite{183foldseek} is a critical tool for rapid protein structure comparison and similarity searching. Foldseek provides an alphabet for describing tertiary structural interactions, consisting of 20 letters, each representing the geometric conformation of each residue and its spatially closest residue (this number intentionally matches the count of natural amino acids). These structural information-containing letters can be used for fast structural searches and to provide structural information to sequence models for building sequence-structure multimodal models.

\subsubsection{Others}

\subsubsection*{\bf UMAP\&t-SNE}

UMAP(Uniform Manifold Approximation and Projection)\cite{184umap} and t-SNE(t-distributed Stochastic Neighbor Embedding)\cite{185tsne} are two popular dimensionality reduction techniques used to embed high-dimensional datasets into lower-dimensional spaces (typically two dimensions) for easier visualization and analysis. They are commonly employed to visualize how PLMs learn from different sequence data. t-SNE is an earlier dimensionality reduction technique that maps the similarities of high-dimensional data into lower-dimensional space, proving more effective when exploring the local structure of data. UMAP, a relatively novel dimensionality reduction method, optimizes local connectivity to preserve global structure, making it more suitable for handling large-scale data.

\section{Discussion}
\label{Discussion}

\subsection{Challenges}

Several practical factors hinder the development of new protein language models. Ambiguous design criteria result in high development costs, and it remains unclear under what model architecture, dataset size, and distribution one model may outperform another\cite{42protflash}. Theoretically, designing larger models should yield better performance under the upper limits of large natural protein databases, and past experiences have confirmed this, leading to substantial resources being allocated to increasing data volume. Additionally, ultra-large models often struggle to generalize to other downstream tasks, necessitating efficient model architecture design and simplification for fine-tuning and alternative methods. While some studies suggest placing greater emphasis on data quality rather than model size\cite{108}, however, this approach heavily relies on high-quality annotations and biological knowledge, which cannot replace the strategy of simply expanding data volume.

The length of protein sequences presents a significant challenge for these models. With protein sequences ranging from approximately 30 to 33,000 amino acids, they are much longer than human language, placing high demands on hardware conditions and often requiring truncation. However, fragmented protein sequences can lead to substantial semantic loss. Although methods such as linear attention mechanisms and stage-wise concatenation operations have been proposed to alleviate hardware pressure, long-sequence data still perform poorly in downstream tasks\cite{42protflash}.

\subsection{Future Directions}

\subsubsection{MSA-free Models}

The use of multiple sequence alignments (MSA) is an unavoidable topic in the protein field. Experimental results indicate that the introduction of MSAs significantly enhances protein language models (PLMs)' performance in downstream tasks \cite{128bfd,161}. However, MSAs also come with notable drawbacks. Firstly, the process of generating MSAs incurs substantial computational costs. Secondly, the impact of MSAs on task outcomes is highly unstable; predictions can vary greatly with different MSAs\cite{186}. More critically, MSA-based models suffer significant performance declines when MSAs are unavailable\cite{59aminobert,161}. Consequently, more and more MSA-free models have been proposed in recent years\cite{161}. These PLMs can mitigate some of the information loss associated with MSAs by training on large-scale data\cite{47esm2,163}. Although MSA-free models still lag behind MSA-based models in performance, their lower costs and broader applicability provide new directions for the development of PLMs.

\subsubsection{Multimodal Models}

Understanding the relationship between sequence, structure, and function is a core issue in the protein field. While the essence of proteins is determined by their sequences, incorporating structural and functional information can provide additional insights to enhance the representation capabilities of PLMs. In recent years, multimodal models that integrate sequence, structure, and function have emerged as a mainstream trend in PLMs, particularly sequence-structure joint models, achieving state-of-the-art performance across multiple downstream tasks. This trend is closely linked to the success of structural prediction models like AlphaFold\cite{128bfd,131alphafold3}, which address the scarcity of experimental protein structure data while offering sufficiently high quality for model training. Although multimodal PLMs are still in their infancy, However, we speculate that they can provide new understandings for more general ways of modeling languages.



\section{Acknowledgments}
This work was supported by National Natural Science Foundation of China under Grant 62172172 and the Postdoctoral Fellowship Program of CPSF under Grant Number GZC20240545.

\vfill

\end{document}